\documentclass[aps,prl,twocolumn,showpacs,superscriptaddress,floatfix]{revtex4}

\usepackage{graphicx}
\usepackage{epsfig}
\usepackage[english]{babel}
\usepackage{amsmath,amssymb}

\newcommand{\be}{\begin{equation}}
\newcommand{\bea}{\begin{eqnarray}}
\newcommand{\eea}{\end{eqnarray}}
\newcommand{\ee}{\end{equation}}

\def\one{\ensuremath{\hbox{$\mathrm I$\kern-.6em$\mathrm 1$}}}
\begin{document}

\title{Renormalization group transformations on quantum states}

\author{F. \surname{Verstraete}}
\affiliation{Max-Planck-Institut f\"ur Quantenoptik,
Hans-Kopfermann-Str. 1,
  Garching, D-85748, Germany.}
\author{J. I. \surname{Cirac}}
\affiliation{Max-Planck-Institut f\"ur Quantenoptik,
Hans-Kopfermann-Str. 1,
  Garching, D-85748, Germany.}
\author{J. I. \surname{ Latorre}}
\affiliation{Dept. d'Estructura i Constituents de la Mat\`eria,
Univ. Barcelona, 08028, Barcelona, Spain.}
\author{E. \surname{Rico}}
\affiliation{Dept. d'Estructura i Constituents de la Mat\`eria, Univ. Barcelona,
08028, Barcelona, Spain.}
\author{M. M. \surname{Wolf}}
\affiliation{Max-Planck-Institut f\"ur Quantenoptik,
Hans-Kopfermann-Str. 1,
  Garching, D-85748, Germany.}

%\date{\today}

\begin{abstract}
We construct a general renormalization group transformation on
quantum states, independent of any Hamiltonian dynamics of the
system. We illustrate this procedure for translational invariant
matrix product states in one dimension and show that product, GHZ,
W and domain wall states are special cases of an emerging
classification of the fixed points of this coarse--graining
transformation.
%We argue that this approach offers a basic way to
%discuss renormalization group irreversibility.

\end{abstract}


\maketitle

The Renormalization Group (RG) provides a procedure to obtain an
effective long distance description of a physical system.
Following Wilson's seminal ideas \cite{Wilson}, RG transformations
are usually constructed in the space of Hamiltonians and are made
of two distinct steps: first, a coarse--graining transformation is
implemented to integrate out short-distance information and,
second, a rescaling of length scales and operators restores the
original picture. This transformation is exact in the sense that
long-distance observables remain unaltered, since they can be
computed either with the original operators and Hamiltonian or
with their renormalized counterparts to yield the same result. The
exact RG transformation can be conveniently truncated so as to
have a very powerful technique to retain only relevant
long-distance degrees of freedom.

The success of RG is ubiquous. Wilson's original idea has been
modified such as to yield the Density Matrix Renormalization Group
(DMRG) \cite{DMRG} algorithm which optimizes the RG truncation;
that is, the choice of relevant degrees of freedom to be retained.
DMRG has been highly successful in describing the ground state
properties in one dimensional non-critical systems, and it can be
understood as a variational method within the set of so-called
matrix product states \cite{AKLT,Fannes} discussed below.

>From a Quantum Information (QI) perspective, renewed attention has
been placed  on quantum  states. Many unexpected quantum state
properties have been discovered and analyzed irrespectively of the
dynamics that may produce them. It is then natural to review our
understanding of RG focusing only on quantum states. This is
actually the very origin of Kadanoff's classical block spin
transformation \cite{Kadanoff}, which has not been pursued on
quantum states so far. The reason for the lack of a quantum
coarse--graining analysis is related to the difficulty of
parametrizing the Hilbert space of many-body systems as opposed to
simple Hamiltonians and to the complexity of dealing with wave
functions in quantum field theory.

\begin{figure}

\begin{picture}(200,80)(0,0)
% first quantum wire
\put(10,55){$H$} \put(30,58){\vector(1,0){60}} \put(110,55){$H'$}
\put(130,58){\vector(1,0){60}} \put(210,55){$H''$}
\put(50,65){WRG}\put(150,65){WRG} \put(0,15){$\{\psi_0\}$}
\put(30,18){\vector(1,0){60}} \put(100,15){$\{\psi_0'\}$}
\put(130,18){\vector(1,0){60}} \put(200,15){$\{\psi_0''\}$}
\put(50,25){qsRG}\put(150,25){qsRG}
 \put(14,50){\vector(0,-1){20}}
 \put(114,50){\vector(0,-1){20}}
 \put(214,50){\vector(0,-1){20}}
\end{picture}
\caption[]{RG evolved states can either be computed by applying
Wilsonian RG to Hamiltonians and then computing their ground
states or by constructing quantum state RG transformations.}
\label{fig:qsflow}

\end{figure}
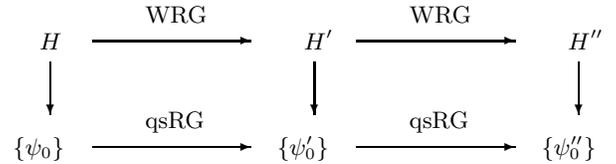

Let us now introduce the general idea of the RG transformation on
pure quantum states for $m$--party systems, where every local
degree of freedom corresponds to a $d$-dimensional Hilbert space.
In analogy with standard RG, we proceed in steps in which we: (i)
merge groups of neighboring particles into new ones, and rescale
the variables correspondingly; (ii) identify states which are
equivalent under local unitary operations. This identification is
motivated by the fact that physics at long scales does not depend
on the choice of local basis and, as it will become clear below,
gives rise to coarse--graining and irreversibility. Technically,
(ii) is realized by introducing an equivalence relation in Hilbert
space, namely $\psi \cong \tilde\psi$ if $\exists U_1, \dots, U_m$
such that $\vert\tilde\psi\rangle=U_1\otimes\dots\otimes
U_m\vert\psi\rangle$, where $U_i$ are local unitary
transformations; that is, two states are equivalent if they differ
by a change in the local basis. Thus, the RG transformation in
each step can be viewed as a map between the resulting equivalence
classes $\{\psi_0\}$ (Fig.\ 1). In practice, we perform the RG
transformation on a representative which is conveniently selected
after each step.

We describe now the above procedure in more detail for a 1D system
with translational symmetry. Given the representative of a class
$\psi_0$, we determine the representative of the class in the next
step, $\psi_0'$, as follows. We pairwise group the sites in the
system and define a coarse--graining transformation for every pair
of local basis states, {\sl e.g.} for the sites $2j$ and $2j+1$,
as $\vert p\rangle_{2j}\vert q\rangle_{2j+1}= \vert pq\rangle_j $.
This transformation yields $\psi_0\to \psi$. Then we have
$\psi_0'=U\otimes \ldots \otimes U |\psi\rangle$, where the
$d^2\times d^2$ unitary matrix $U$ performs the change of
representative in the coarse--grained space. Let us emphasize that
the freedom to take a unitary transformation in the
coarse--grained spaced goes beyond the onset freedom made by the
product of unitary matrices in the original Hilbert spaces. The
matrix $U$ can be non-local as seen by the $2j$ and $2j+1$ sites.
Some local information is now washed out, while preserving all the
quantum correlations relating the coarse--grained block to other
ones.

Operators also get coarse--grained along the above transformation.
Take for instance an operator acting on one local Hilbert space,
{\sl e.g.} $O_{2j}$. Expectation values must remain unchanged,
 \be
 \langle \psi_0\vert O_{2j}\vert \psi_0\rangle
 = \langle \psi_0'\vert O'_{j}\vert
 \psi_0'\rangle
 \ee
which leads to
 \be O'_l=U (O_{2j}\otimes \one_{2j+1})U^{\dagger}
 \ee
where $\one$ is the identity matrix. To complete a RG
transformation we simply need to rescale distances, i.e., to
double the lattice spacing.

Exact RG transformations are often truncated in order to become of
practical use. Let us, for example, consider the state after $n$
coarse--graining steps, where each \emph{local} site corresponds
to $2^n$ original spins. We are interested in describing
long--range effects, hence in the degrees of freedom of the
$d^{2^n}$ dimensional Hilbert space that couple to the outer sites
(all the other information is local). In the case of ground states
of 1-D noncritical spin chains for example, we know that the
entropy of a block of spins saturates to a finite value,
indicating that indeed very few degrees of freedom couple to the
outer sites \cite{Vidal}. Mathematically, this means that there
exists a local isometry $U$ that transforms the $d^{2^n}$
dimensional space into a much smaller one, without practically
affecting the correlations between the different blocks. The whole
issue of truncation in RG then consists of coming up with an
optimal algorithm for keeping the relevant degrees of freedom.
Obviously, these relevant degrees of freedom will exactly be the
ones that correspond to the largest weights in the reduced density
operator of the block of $2^n$ spins. This is very much related to
the concept of the density operator in DMRG, although in that case
only half-infinite blocks are considered.

In the following we explicitly carry out the RG transformations in
terms of matrix product states (MPS). In this representation, the
whole procedure can be naturally implemented, and the nature of
the fixed points becomes more transparent. Any 1D translationally
invariant state can be written in its MPS form as \cite{Fannes}
 \be \vert
 \psi\rangle=\sum_{p_1,p_2\dots,p_m=1}^d {\rm Tr} \left(
 A^{p_1} A^{p_2} \dots A^{p_m}
 \right) \vert p_1, p_2,\dots,p_m\rangle
 \ee
where the $D\times D$ matrices $\{A^p\}$ parameterize the state.
The value of the dimension $D\le d^{m/2}$ depends on the
particular state \cite{VPC04}.

\begin{figure}[t]
  %\centering
  \includegraphics[width=1.0\linewidth]{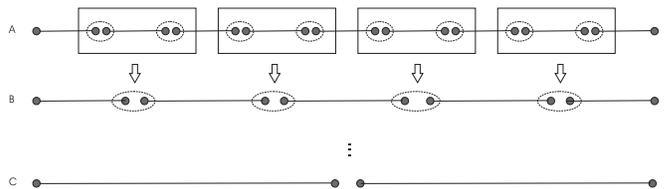}
  \caption{Coarse-graining matrix product states}
  \label{Fig2w}
\end{figure}

The quantum coarse--graining procedure where we map two
neighboring spins to one new block spin can be fully characterized
in terms of the matrices $A$ for the corresponding
representatives. These matrices can be conveniently chosen
starting from the coarse--grained matrices $\tilde A^{(pq)}:= A^p
A^q$ through the singular value decomposition
 \be
 \label{svd}
 (\tilde A^{(pq)})_{\alpha\gamma}= \sum_{l=1}^{\min(d^2,D^2)}
 (U^\dagger)^{(pq)}_l \lambda^l (V^l)_{\alpha\gamma}.
 \ee
>From this decomposition we identify the isometry $U$ which selects
the representative and the coarse--grained tensor $A'$
 \be
 A^p\stackrel{RG}{\longrightarrow}
 A^l =\lambda^l V^l.
 \ee
The advantage of this representative is that the Hilbert space
corresponding to the block spins remains bounded above by $D^2$ at
any step, as it is clear from the decomposition (\ref{svd}). That
is, by coarse--graining we do not have to increase the dimension
of the spins once we reach $D^2$, and therefore, it is possible to
perform an exact coarse--graining on finite dimensional matrix
product states without any need of truncation! Obviously, the
interesting question is now to classify all possible fixed points
of this exact renormalization flow.

Before introducing a formalism to characterize those fixed points,
let us present some simple examples. The first one is provided by
product states (for which $D=1$, $A^1=1$ and $A^k=0$ for
$k=2,\ldots,d$ and $A'^i=A^i$). These are precisely the state
obtained for massive theories in their infrared fixed points. A
significant further example corresponds to GHZ-like states
\cite{GHZ}. If we take $d=2$ and $A^0=\vert 0\rangle \langle
0\vert,\ A^1=\vert 1 \rangle \langle 1\vert$ the RG transformation
reads $A^i\stackrel{RG}{\longrightarrow} A'^i=A^i$ with
$U=|0\rangle\langle 00|+|1\rangle\langle 11|$. Those states indeed
appear as RG infrared points in {\sl e.g.} the quantum Ising chain
for vanishing external magnetic field \cite{LLRV}.

Let us now construct a general formalism by which the fixed points
can be characterized. In order to circumvent the arbitrariness in
the choice of the local bases, we introduce the auxiliary
$D^2\times D^2$ transfer matrix
 \begin{equation}
 E=\sum_{p=1}^D A^p\otimes \bar A^{p},
 \label{E}
 \end{equation}
where the bar indicates complex conjugation. Note that if we
choose $\tilde A_q = \sum_p U_p^q A^p$ with $U$ unitary, we have
 \be
 \tilde E=\sum_{q=1}^D \tilde A^q\otimes \tilde{\bar A^{q}}
 =\sum_{p=1}^D A^p\otimes \bar A^{p}=E.
 \ee
Conversely, the matrix $E$ uniquely defines the matrices $A^p$ up
to such a local unitary operator, and thus it parameterizes the
equivalence class of D-dimensional MPS where all elements of the
class are related by local unitary operations \cite{note}.
Furthermore, a RG step corresponds to the simple transformation
$E\stackrel{RG}{\longrightarrow} E'=E^2$. Therefore, in order to
study the alluded fixed points we just have to characterize the
class of possible operators
$\{E_{\infty}\}=\{\lim_{n\rightarrow\infty}E^n\}$ for some $E$ in
the form (\ref{E}). Since we can always choose the largest
eigenvalue of $E$ to be equal to 1 \cite{note1}, we just have to
characterize all matrices $E$ of the form (\ref{E}) that have only
eigenvalues of magnitude $1$ or $0$ (eigenvalues smaller than 1
will decay exponentially along repeated coarse--graining steps).

In the generic case, the largest eigenvalue of $E$ will be
non-degenerate and both its left and right eigenvectors will have
maximal Schmidt rank; that is, $E^\infty=|\Phi_R\rangle\langle
\Phi_L|$, where the reduced density operators of $\Phi_{L,R}$ are
rank $D$ matrices. With the similarity transformation \cite{note1}
we can always choose $|\Phi_R\rangle=\sum_{i=1}^D|ii\rangle$ and
$|\Phi_L\rangle=\sum_{i=1}^D\lambda_i|ii\rangle$, with
$\lambda_i>0$. From these eigenvectors we can directly read off
the matrices $A^{(pq)}=\sqrt{\lambda_q} |p\rangle\langle q|$ with
$p,q=1,\ldots,D$. Thus, at the fixed point, the scale--invariant
state (representative of the equivalence class) can be written in
terms of two auxiliary spins at each point, which are in an
entangled state $|\Phi_L\rangle$ with its neighbors as shown in
Fig.\ 2(c). This conclusion is very appealing, and has the
following consequences:

\begin{itemize}

 \item All connected correlation functions of the form $\langle
O^iO^j\rangle-\langle O^i\rangle\langle O^j\rangle$ are exactly
zero when $j>i+1$; this is exactly what one expects from a generic
state by looking at it in a coarse--grained way: the correlations
decay exponentially along the renormalization flow and become zero
at the fixed point.

 \item The entropy of a block of spins of length $L$ is
 independent of $L$ and is exactly twice the entropy of
 entanglement of $|\Phi_R\rangle$.

 \item In terms of the picture introduced in \cite{VPC04},
 coarse--graining turns the virtual underlying spins into real ones but
 changes the maximally entangled states into $|\Phi_R\rangle$.

\end{itemize}

The qualitative features of the renormalization flow can also be
easily understood. For example, an observable that only acts
nontrivially on the spins in the center of a block converges
exponentially fast to the identity operator times its expectation
value; this was indeed expected as observables defined in the
middle of a block of length $L$ with $L$ larger than the
correlation length should not be able to act nontrivially on the
spins far away. On the other hand, the entropy of a block
increases by coarse--graining (if we do not rescale the lengths).
By the strong subadditivity of the entropy
(cf.\cite{NielseChuang}) this is even true in general for all
translationally invariant states.

As an illustration of the generic case analyzed above, let us
consider the ground state of the AKLT Hamiltonian \cite{AKLT}, for
which $A^p=\sigma^p$ ($p=1..3$) are the Pauli matrices. Some
simple algebra shows that $E$ has eigenvalues $3,1,-1,-1$ and
hence no degeneracy in the largest eigenvalue. The fixed point is
given by the eigenvector corresponding to eigenvalue $3$, namely
$E_i=|I\rangle\langle I|$ with $|I\rangle=\sum_{i=1}^2|ii\rangle$.
Hence, it is a dimer of maximally entangled states, i.e., in the
language of QI a perfect resource for a quantum repeater.
Intriguingly, this is also the exact fixed point of the 1-D
cluster state \cite{cluster} represented by $A^0=\vert 0 \rangle
(\langle 0|+\langle 1|)$ , $A^1=\vert 1 \rangle (\langle
0|-\langle 1|)$. However, in this case the fixed point is already
obtained after one coarse--graining step as the corresponding $E$
contains a nilpotent Jordan block:
 \be
 E_=\frac{1}{2}\left(\begin{array}{cccc}
 1&1&1&1\\0&0&0&0\\0&0&0&0\\1&-1&-1&1\end{array}\right)\hspace{.5cm}
 E_{cl}^2=\sum_{ij}|ii\rangle\langle jj|.
 \ee

The formalism developed above can also be applied in the case
$D\to \infty$. To illustrate that, let us consider an Ising chain
with transverse magnetic field or an anisotropic Heisenberg
antiferromagnet. In the non--critical regime, their correlation
length $\xi$ is finite, which implies that the associated operator
$E$ will have a non-degenerate largest eigenvalue ($1/\xi\propto
\ln(\lambda_1/\lambda_2)$). Accordingly, the fixed point of the
coarse--graining map will therefore consist of a dimer state. The
eigenvalues of the reduced density operator of a half-infinite
chain are precisely the Schmidt coefficients of the bipartite
states making up the dimer. We can in fact determine these
coefficients by using the results of Peschel and col.
\cite{Peschel}, who have been able to determine the mentioned
reduced density operator exactly. For the Ising model we have
\be
\lambda_{(n_1,n_2,\cdots
 n_\infty)}=
\begin{cases}
e^{-\epsilon\sum_{j=0}^{\infty}(2j+1)n_j} & \lambda<1\\
e^{-\epsilon\sum_{j=0}^{\infty}2jn_j} & \lambda>1.
\end{cases}
\ee
Here $\epsilon=\pi K[(1-\mu^2)^{1/2}]/K(\mu)$, with $K$ the
complete elliptic integral of the 1st kind,
$\mu=\min(\lambda,1/\lambda)$, and $n_j\in\{0,1\}$. In the case of
the noncritical Heisenberg  model $H_{XXZ}=\sum_i
\sigma^x_i\sigma^x_{i+1}+ \sigma^x_i\sigma^x_{i+1}+
\Delta\sigma^z_i\sigma^z_{i+1}$ with $\Delta>1$, the eigenvalues
are as in the Ising case with $\lambda>1$, but with $\epsilon={\rm
arccosh}(\Delta)$.

We consider now the full classification of fixed points for the
simplest non--trivial case $D=2$. By considering the right
eigenvectors of $E$ corresponding to the maximal eigenvalue, we
have two possibilities \cite{note2}: (a) one of them, say
$|\Phi_R\rangle$, is entangled; (b) there exists only one
eigenvector which corresponds to a product state.

In the first case (a), by using the similarity transformation
\cite{note} we can always take $|\Phi_R\rangle=\sum_{i=1}^2
|ii\rangle$. Using the isomorphism \cite{note2} and the full
classification of trace preserving completely positive maps acting
on a qubit \cite{Ruskai} one can prove that $E_\infty$ has rank
$1,4$ or $2$. For  the first cases, we recover the generic case
studied above or obtain a product state, respectively. When the
rank is 2 we obtain $A^0=|0\rangle\langle 00|;A^1=|1\rangle\langle
11|$, i.e., the GHZ state \cite{GHZ} studied above.

The situation is more complicated in the second case (b), as this
implies that $E$ is not diagonalizable but has a Jordan--block
decomposition. Some tedious but straightforward algebra leads to
two different possibilities for the fixed points. In the first
case the effective Hilbert space is 2-dimensional and
 \be
 A^0=\begin{pmatrix}1&0\\0&\exp(-i\theta)\end{pmatrix}
 \hspace{2cm}
 A^1= \begin{pmatrix}0&0\\1&0\end{pmatrix}.
 \ee
Note that $A^0A^1=e^{-i\theta}A^1$ and that $A^1$ is nilpotent;
this immediately implies that the state, when written in the
computational basis, consists of terms like $|00\cdots 010\cdots
0\rangle$ where at most one $|1\rangle$ appears. When $\theta=0$
we recover the well--known $W$-state \cite{W}, which is indeed
scale invariant. In the second case of Jordan fixed points, the
spins have effective support on a Hilbert space of dimension 3
(but 2 when $\cos(\alpha)=1$), and a possible decomposition is
given by
\be
\begin{split}
A^0=&\begin{pmatrix}0&0\\\cos(\alpha)\sin(\beta) &
e^{i\theta}\end{pmatrix} 
\hspace{.5cm}
A^1=\begin{pmatrix}0&0\\ \sin(\alpha)&0 \end{pmatrix} \\
&A^2=\begin{pmatrix}e^{-i\theta}&0\\ \cos(\alpha)\cos(\beta)&0 \end{pmatrix}.
\end{split}
\ee
As $A^0A^1\propto A^1,A^1A^2\propto A^2,A^0A^2\propto
A^1,A^1A^0=0,A^2A^1=0,A^2A^0=0$, the state will be a superposition
of terms of the form $|00\cdots 0122\cdots 2\rangle$. Therefore
these scale-invariant states represent linear combination of
domain walls.

For the case $D=2$ this completes the classification of all fixed
points, which remarkably correspond to almost all well-studied
multipartite states encountered in QI. We note that in the
thermodynamic limit case (b) in the classification is redundant in
the sense that W and product state as well as domain wall and GHZ
state become then locally indistinguishable. A distinction can,
however, be relevant from a QI perspective, where it is more
natural to carry out only a finite number of coarse--graining
steps which enlarge the region of local accessibility only to a
physically reasonable vicinity. In this context the system to be
transformed can thus be finite.

 For $D>2$ a coarse classification of the fixed
points can be given by their possible decompositions into ergodic
states and their periodic components discussed in \cite{Fannes}.

Finally, let us mention that we have concentrated here on 1D
systems, since in that case MPS give a very useful description.
Fortunately, the analogue of MPS in higher dimensions has been
recently put forward \cite{PEPS}, which may allow us to extend the
analysis  introduced here to two and three spatial dimensions.

Let us conclude with a comment on irreversibility of RG flows.
Our RG transformation defined on finite matrix product
states incorporates unitarity in a natural way at variance
with other approaches.
Moreover, the eigenvalues of $E^2$ are smaller than
the ones of $E$. RG flows will inevitably eliminate all
the eigenvalues smaller than 1, which can
be phrased as irrelevant pieces of the state.
A careful treatment of irreversibility of RG flows should
incorporate the infinite dimensional case, associated to conformal
field theories, and the discussion of degeneracy in the
final relevant Hilbert space.

Work supported by DFG (SFB 631), European projects (RESQ,
TOPQIP, CONQUEST), Kompetenznetzwerk der Bayerischen
Staatsregierung Quanteninformation and MCyT.


\begin{thebibliography}{99}
\bibitem{note1} According to \cite{note}, $E$ and $|c|^2(X\otimes \bar
X)E(X\otimes \bar X)^{-1}$ provide the same description of the
state. We can use this fact to demand certain properties of $E$.
\bibitem{note2} Matrices $E=\sum_p
A^p \otimes \bar A^p$ can be isomorphically mapped into completely
positive maps, $\$_E$, which act as follows: $\$_E(\rho)=\sum_p
A^p \rho A^{\dagger p}$. Thus, we can use the knowledge of the
classification of those maps \cite{Ruskai} in order to classify
the structure of $E$.
\end{thebibliography}
\end{document}